\documentclass[aps,prx,amsmath,amssymb,twocolumn,superscriptaddress,showpacs,longbibliography]{revtex4-1}
\usepackage{amsmath,amssymb}
\usepackage{graphicx}
\usepackage{appendix}
\usepackage[usenames,dvipsnames]{color}
\usepackage[colorlinks=true, citecolor=blue, linkcolor=blue, urlcolor=blue]{hyperref}
\usepackage{bm}

\usepackage[nottoc,numbib]{tocbibind}

\newcommand{\D}{\textrm{d}}
\newcommand{\I}{\textrm{i}}
\newcommand{\ket}[1]{\left|#1\right\rangle}

\begin{document}

\title{Topological Floquet Engineering of Twisted Bilayer Graphene}

\author{Gabriel E.~Topp}
\affiliation 
{Max Planck Institute for the Structure and Dynamics of Matter, Luruper Chaussee 149, 22761 Hamburg, Germany}

\author{Gregor Jotzu}
\affiliation 
{Max Planck Institute for the Structure and Dynamics of Matter, Luruper Chaussee 149, 22761 Hamburg, Germany}

\author{James W.~McIver}
\affiliation 
{Max Planck Institute for the Structure and Dynamics of Matter, Luruper Chaussee 149, 22761 Hamburg, Germany}

\author{Lede Xian}
\affiliation 
{Max Planck Institute for the Structure and Dynamics of Matter, Luruper Chaussee 149, 22761 Hamburg, Germany}

\author{Angel Rubio}
\affiliation 
{Max Planck Institute for the Structure and Dynamics of Matter, Luruper Chaussee 149, 22761 Hamburg, Germany}
\affiliation{Center for Computational Quantum Physics (CCQ), The Flatiron Institute, 162 Fifth Avenue, New York NY 10010}

\author{Michael A.~Sentef}
\email{michael.sentef@mpsd.mpg.de}
\affiliation 
{Max Planck Institute for the Structure and Dynamics of Matter, Luruper Chaussee 149, 22761 Hamburg, Germany}

\date{\today}

\begin{abstract}
We investigate the topological properties of Floquet-engineered twisted bilayer graphene above the magic angle driven by circularly polarized laser pulses. Employing a full Moir\'e-unit-cell tight-binding Hamiltonian based on first-principles electronic structure we show that the band topology in the bilayer, at twisting angles above 1.05$^\circ$, essentially corresponds to the one of single-layer graphene. However, the ability to open topologically trivial gaps in this system by a bias voltage between the layers enables the full topological phase diagram to be explored, which is not possible in single-layer graphene. Circularly polarized light induces a transition to a topologically nontrivial Floquet band structure with the Berry curvature of a Chern insulator. Importantly, the twisting allows for tuning electronic energy scales, which implies that the electronic bandwidth can be tailored to match realistic driving frequencies in the ultraviolet or mid-infrared photon-energy regimes. This implies that Moir\'e superlattices are an ideal playground for combining twistronics, Floquet engineering, and strongly interacting regimes out of thermal equilibrium. 
\end{abstract}
\maketitle

\section{Introduction}\label{INTRO}

Light-matter coupled systems are emerging as an important research frontier bridging condensed matter physics \cite{basov_towards_2017}, quantum optics \cite{ebbesen_hybrid_2016, flick_atoms_2017, flick_strong_2018, ruggenthaler_quantum-electrodynamical_2018,schafer_modification_2019}, as well as cold atoms in optical lattices \cite{hemmerich_effective_2010, bukov_universal_2015,eckardt_colloquium:_2017}. In ultrafast materials science, intriguing phenomena have been explored, including but not limited to ultrafast switching between different phases of matter \cite{rini_control_2007, stojchevska_ultrafast_2014, claassen_universal_2019}, light control of important couplings in solids \cite{singla_thz-frequency_2015, pomarico_enhanced_2017, tancogne-dejean_ultrafast_2018}, and light-induced superconductivity \cite{fausti_light-induced_2011, mitrano_possible_2016}. In cavities, spectacular effects have been observed or predicted, such as dramatically enhanced conductivity in polymers \cite{orgiu_conductivity_2015}, cavity-modified materials properties \cite{sentef_cavity_2018, mazza_superradiant_2019, curtis_cavity_2019, kiffner_manipulating_2019}, novel spectroscopies using the quantum nature of light \cite{ruggenthaler_quantum-electrodynamical_2018}, or light-controlled chemical reaction pathways \cite{ribeiro_polariton_2018}. Finally, in optical lattices, periodically driven quantum systems are investigated within the realm of Floquet engineering, in which the driving is used as a tool to generate effective Hamiltonians with tunable interactions \cite{zenesini_coherent_2009,struck_quantum_2011, aidelsburger_experimental_2011, jotzu_experimental_2014, kennedy_observation_2015}, which has also been demonstrated in purely photonic systems \cite{rechtsman_photonic_2013}. 

\begin{figure*}[htp!]
    \centering
    \includegraphics[width=1.0\linewidth]{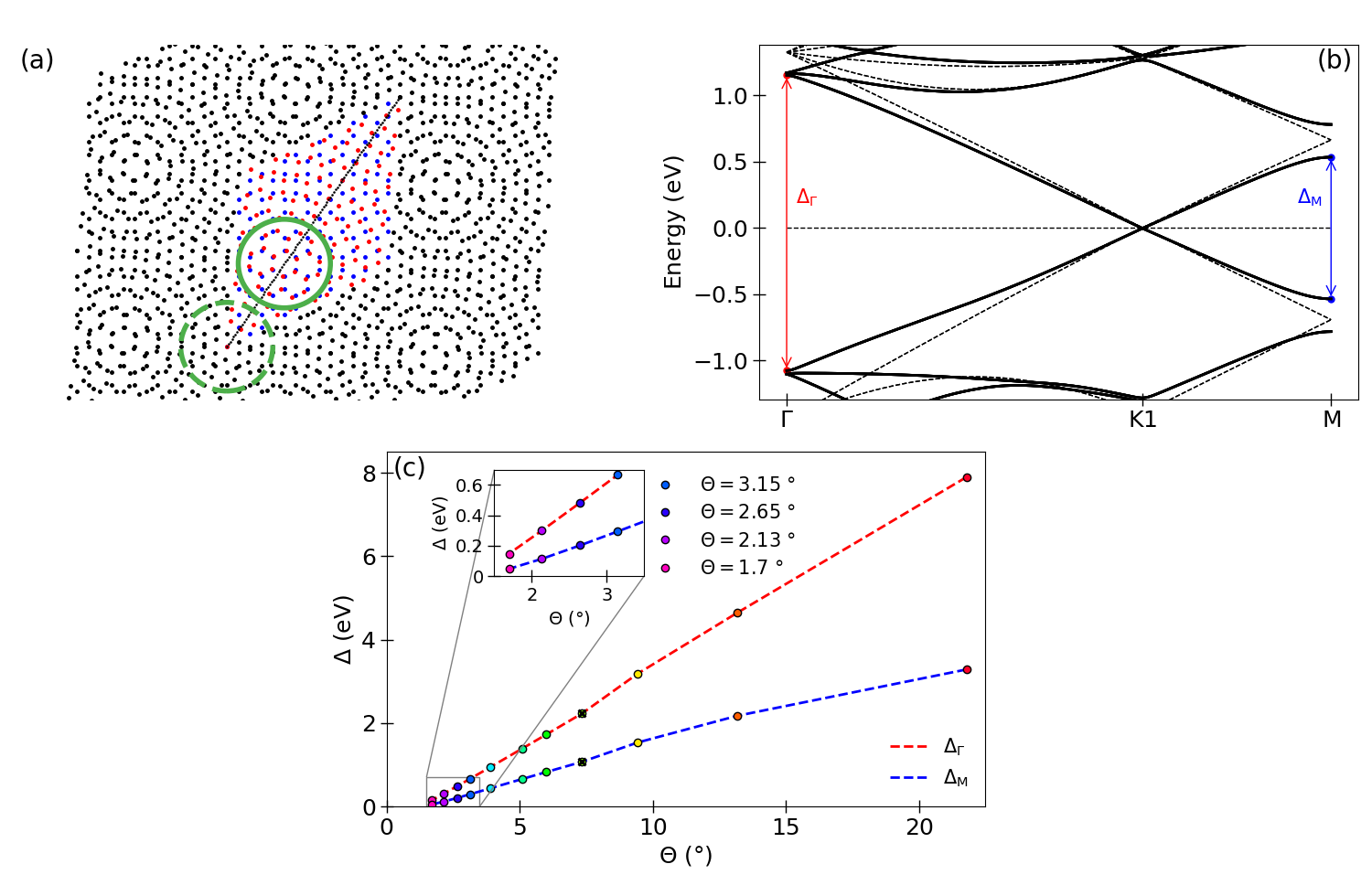}
    \caption{Atomic structure and tight-binding model bands of equilibrium twisted bilayer graphene. (a) Real-space atomic lattice. Red (blue) points denote the carbon atoms of the bottom (top) layer of one supercell. The black dotted line indicates the mirror-symmetry line which maps the in-plane atomic coordinates of both layers onto each other. The origin is chosen on an $AA$-stacked carbon site. The twist angle is ${\Theta = 7.34^\circ}$, which corresponds to a supercell with 244 atoms. The full (dashed) circle indicates a region of $A$-$A$ ($A$-$B$) stacking. (b) Equilibrium band structure along {$\Gamma$-$K_1$-M}, where $|\Gamma-\text{$K_1$}| = 0.22$ $\text{\AA}^{-1}$. The dashed lines show the single-layer band structure without interlayer coupling but folded back into the reduced zone. The bandwidth at $\Gamma$ (M) is indicated by a red (blue) arrow. (c) Equilibrium bandwidth of the valence and conduction bands at $\Gamma$ (red line) and $M$ (blue line) for different twist angles, $\Theta$ ($\Theta = 7.34^\circ$ values colored black). The inset shows small angles, whose associated bandwidths lie within the mid-infrared photon energy range.}
    \label{fig:1} 
\end{figure*}   

In solids, key progress has been made for instance in observing Floquet-Bloch states in time-resolved photoemission spectroscopy \cite{wang_observation_2013} or with the recently demonstrated light-induced anomalous Hall effect in graphene under circularly polarized laser driving \cite{mciver_light-induced_2018, sato_microscopic_2019, sato_light-induced_2019}. However, the concept of Floquet engineering of a material's topological properties \cite{oka_photovoltaic_2009,kitagawa_transport_2011,lindner_floquet_2011,rudner_anomalous_2013,usaj_irradiated_2014,foa_torres_multiterminal_2014,sentef_theory_2015,mikami_brillouin-wigner_2016,dehghani_out--equilibrium_2015,hubener_creating_2017} has been limited to very few material systems so far. The reason for this is that the time and energy scales often cause problems in avoiding resonant excitations and accompanying heating effects that are usually detrimental for the observation of interesting light-induced states of matter, unless full thermalization is delayed \cite{abanin_rigorous_2017, topp_all-optical_2018}, or unless the laser is only used as a means to break the symmetry for transient intermediate states, such as in the case of laser-controlled novel superconductors \cite{claassen_universal_2019}. It is therefore an important task to identify material platforms in which crucial energy scales, such as the relevant effective electronic bandwidth, can be tuned with respect to the photon energy of the pump laser. This is exactly the case in twisted van der Waals heterostructures forming Moir\'e superlattices \cite{bistritzer_moire_2011}. In particular, twisted bilayer graphene (TBG) has recently attracted considerable attention due to the discovery of superconductivity \cite{cao_unconventional_2018} along with correlated insulating phases in its vicinity \cite{cao_correlated_2018}. In TBG, the Fermi velocity and electronic bandwidth of the Dirac bands can be tuned by changing the twisting angle, leading to flat bands at particular magic angles and opening the possibility of \textit{twistronics} \cite{ribeiro-palau_twistable_2018} and tunable energy absorption spectra \cite{moon_energy_2012}.
In addition, this paves the way to tune the role of interactions in a materials setting \cite{wolf_electrically-tunable_2019}, similarly to the case of artificial lattices, in which more exotic Floquet Hamiltonians have been generated \cite{aidelsburger_measuring_2015}. In the context of TBG, this might for instance open the possibility to study strongly correlated Floquet-engineered phases of matter such as the proposed fractional Floquet-Chern insulator \cite{grushin_floquet_2014}.

In this work we perform a full tight-binding model calculation for TBG for an intermediate rotation angle, ${\Theta = 7.34^\circ}$ with a unit cell containing 244 sites, in order to investigate its electronic and topological properties in and out of equilibrium above the magic angle regime in a microscopic picture. A circularly polarized laser field with a frequency of 2.23 eV, tuned to the bandwidth of the lowest energy band manifold, breaks time-reversal symmetry and induces a topologically nontrivial phase, which is tracked by the Berry curvature of Floquet bands projected onto the bare energy eigenstates. We find that the topology of TBG above the magic angles ultimately corresponds to two copies of single-layer graphene. For circularly polarized driving, we find a nontrivial Chern-insulating band structure (with Chern number $C=4$), while in equilibrium, the system exhibits trivial topology with cancellations of valley Berry curvature when inversion symmetry is broken by a backgate bias voltage between the layers. This offers the unique opportunity, in contrast to single-layer graphene, to study the transition between the topologically trivial and non-trivial phases, as originally envisioned by Haldane in his seminal work on the quantum anomalous Hall effect \cite{haldane_model_1988}.
Notably this is unlike the case of time-reversal symmetry breaking by a magnetic field, where the large magnetic unit cell and corresponding small Brillouin zone cause dramatically different effects such as the emergence of a Hofstadter butterfly \cite{ponomarenko_cloning_2013}.
Our results confirm that for angles larger than the highest magic angle (1.05$^{\circ}$) the interlayer interactions can be captured by perturbative treatment, which maintains the linear band dispersions while renormalizing the Fermi velocities, as stated in \cite{trambly_de_laissardiere_localization_2010, trambly_de_laissardiere_numerical_2012}.  Furthermore, we explicitly show that this is not only true for the electronic band energies but also for the system topology. Thus we expect a time-resolved Hall current measurement, in the time-reversal symmetry-broken state, to show a nonzero Hall conductance \cite{haldane_model_1988, oka_photovoltaic_2009} approaching a quantized value of $4e^2/h$ for completely filled valence bands and empty conduction bands, where the factor of $4$ reflects two valence bands, two spin species, and two valleys. 

The paper is organized as follows. In Sec.~\hyperref[Sec.I]{I}, we explain the basic properties of our tight-binding model. In Sec.~\hyperref[Sec.II]{II}, we investigate the electronic properties of our starting point, particularly the equilibrium topology of the system. Sec.~\hyperref[Sec.III]{III} outlines the Floquet electronic band structure and its key differences, compared to equilibrium. In Sec.~\hyperref[Sec.IV]{IV} we present our main result of the Floquet engineered Berry curvature and explore the Floquet-topological phase space in dependence of driving amplitudes and different choices of local potentials. We finish by a brief conclusion in Sec.~\hyperref[CONC]{V}.


\section{Tight-binding model}\label{Sec.I}

Our starting point are two $A$-$A$-stacked graphene layers \footnote{One could equivalently start from $A$-$B$ stacking, and we have checked that all of the results presented in this work hold equally for both choices.}. Choosing the origin of the cartesian coordinate system at an atomic site, we construct the bilayer by a rotation of the top layer by a twist angle, $\Theta = 7.34^\circ$. This rotation amounts to a reflection of the bottom layer within the $x$-$y$-plane along the supercell diagonal, as depicted in Fig.~\ref{fig:1}(a). Using the commensurability condition derived in \cite{trambly_de_laissardiere_localization_2010}, this results in a supercell of 244 carbon atoms, which has regions of $A$-$A$- and $A$-$B$-stacking, respectively. 
We use a general tight-binding hopping Hamiltonian of the form
\begin{equation}
    H = \sum_{i}\epsilon_{i}c_{i}^\dagger c_{i} + \sum_{i\neq j}t_{ij} c_{i}^\dagger  c_{j}, 
\end{equation}
where $c_{i}^\dagger$ ($c_{i}$) creates (annihilates) an electron in the $p_z$ orbital at the atomic position $\bm{r}_i$. The variables $\epsilon_{i}$ and $t_{ij}$ denote the local potential and hopping matrix elements, respectively. The model parameters are given in Appendix \ref{app:model}.

The equilibrium band structure along the symmetry path {$\Gamma$-$K_1$-$M$} in the mini-Brillouin zone (mBZ) of the Moir\'e superlattice is presented in Fig.~\ref{fig:1}(b). It shows a linear Dirac-cone dispersion around one Dirac point $K_1$ with a renormalized Fermi velocity, $v_F/v_F^0=0.9$, compared to the one in single-layer graphene (dashed lines). The renormalized $\Gamma$-point bandwidth $\Delta_\Gamma$ and $M$-point energy difference $\Delta_M$ for the bottom conduction and top valence bands are indicated by colored arrows. 

Envisioning periodic laser driving at specific photon energies with the goal to Floquet-engineer a light-induced anomalous Hall effect \cite{mciver_light-induced_2018, sato_microscopic_2019}, we now address the relevant electronic energy scales as a function of twisting angle.
To this end we show in Fig.~\ref{fig:1}(c) both $\Delta_\Gamma$ and $\Delta_M$ for different commensurate rotation angles above the magic-angle regime. As discussed in various references, such as \onlinecite{bistritzer_moire_2011}, the low-energy Dirac bands can be rescaled to exhibit a desired bandwidth by tuning to particular twist angles. In the inset of Fig.~\ref{fig:1}(c) we focus on the range of angles for which the $\Delta$ values lie in the range attainable in an existing ultrafast transport setup \cite{mciver_light-induced_2018}. This opens up the possibility of a Floquet-engineered band structure with a light-induced anomalous Hall effect if the laser breaks time-reversal symmetry (see below) with electronic energy scales tunable near the photon energy. This might for instance be possible for a twist angle of 1.7$^\circ$ and photon energy of 200 meV, which matches roughly the low-energy bandwidth in this case. 

\section{Equilibrium Topology}\label{Sec.II}

\begin{figure*}[htp!]
    \centering
    \includegraphics[width=1.0\linewidth]{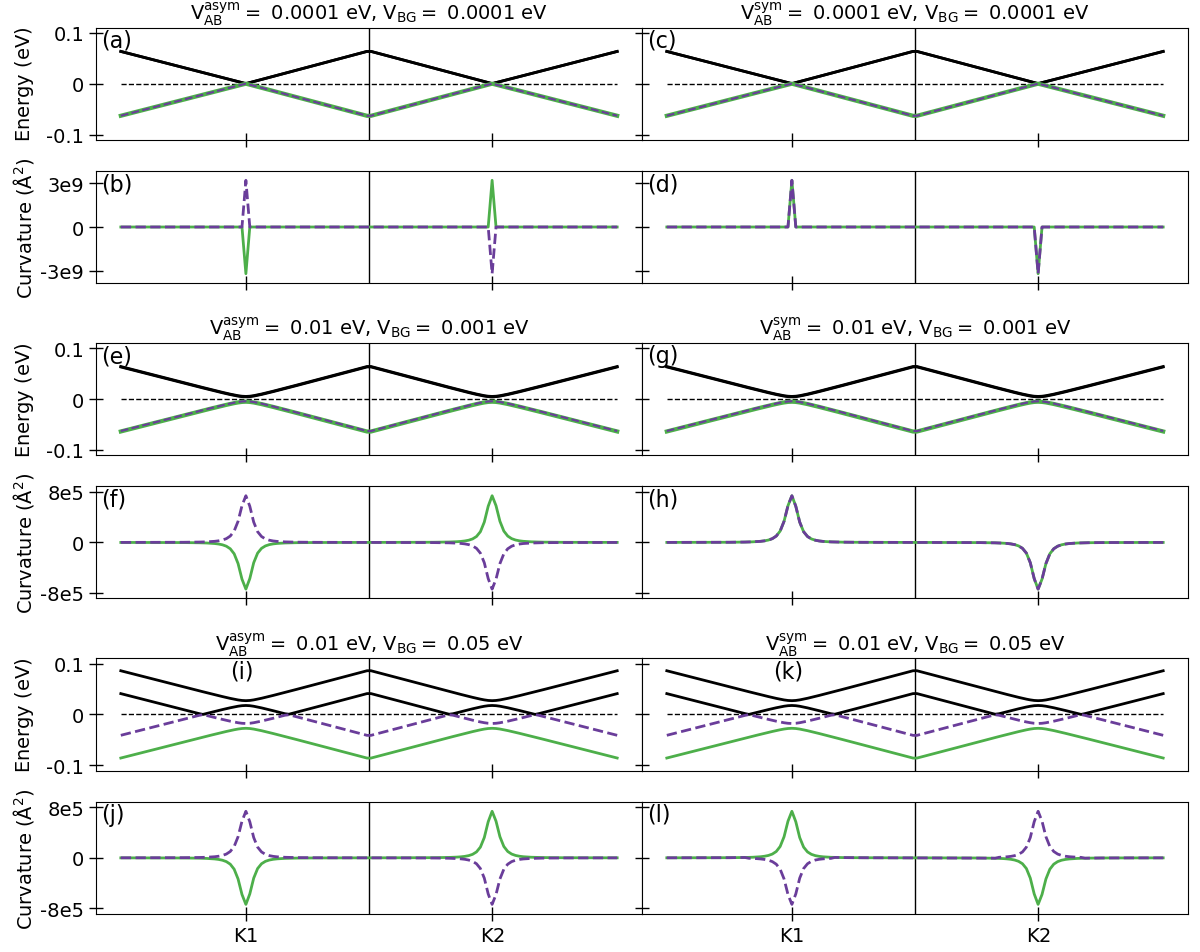}
    \caption{Equilibrium band structure and corresponding valence-band Berry curvatures in close proximity to the Dirac points of the mBZ along {$\Gamma$-$K_1$-$M$-$K_2$-$\Gamma$} (interval of length 0.022 $\text{\AA}^{-1}$ around $K_1$ and $K_2$) for different onsite potentials $V_{AB}$ and backgate voltages $V_{BG}$. (a),(c),(e) Antisymmetric $\pm/\mp$ sublattice potential within the layers. (g),(i),(l) Symmetric $\pm/\pm$ sublattice potentials. (b),(f),(j),(l) Local and valley band curvature shows opposite sign, respectively. (d),(h) Non-vanishing valley Berry curvature but zero integrated Berry curvature due to sign flip between valleys.}
    \label{fig:2} 
\end{figure*}  

Before turning to Floquet-engineered bands and topology, we first discuss the equilibrium Berry curvature near the Dirac points when different onsite potentials are applied to TBG. This can be seen as a direct generalization of the Haldane model for single-layer graphene \cite{haldane_model_1988} with broken inversion symmetry and intact time-reversal symmetry. 

For the onsite potentials we choose from three different options, detailed in Appendix \ref{app:model}: (i) a backgate voltage which corresponds to an energetic difference $V_{BG}$ between the top and bottom layers; (ii) an asymmetric $A$-$B$ potential with energetic difference $V^{\text{asym}}_{AB}$ between $A$ and $B$ sublattice in one layer and $-V^{\text{asym}}_{AB}$ in the other layer; and (iii) a symmetric $A$-$B$ potential with energetic difference $V^{\text{sym}}_{AB}$ between $A$ and $B$ sublattice in both layers. As we will see, different combinations of these potentials (i) and (ii) and (i) and (iii), respectively, lead to different band structures and different Berry curvatures within the bands. We notice that for the real TBG device, $V^{\text{sym}}_{AB}=0$ and $V^{\text{asym}}_{AB}=0$. These onsite potentials should be seen as fictitious potentials that illustrate how the topology changes for different ways in which inversion symmetry can be broken, similarly in spirit to the Haldane model. However, the backgate potential can be applied experimentally, as has been demonstrated for conventional $A$-$B$-stacked bilayer graphene \cite{zhang_direct_2009,taychatanapat_electronic_2010}. Unlike the case of $A$-$B$-stacked bilayer graphene, which has a parabolic energy dispersion, TBG does have a Dirac-like dispersion, which offers the unique opportunity to experimentally study the competition between broken time-reversal and broken inversion symmetries.

Fig.~\ref{fig:2} shows the evolution of band structures and Berry curvatures (see Appendix \ref{app:curvature}) near both Dirac points $K_1$ and $K_2$ in the mBZ for twist angle ${\Theta = 7.34^\circ}$. First we focus on infinitesimally small $V^{\text{asym}}_{AB}$ $=$ $V_{BG}$ $=$ 0.0001 eV (Fig.~\ref{fig:2}(a),(b)). Both the valence and conduction bands are still almost twofold degenerate and a small energy gap of order 0.0001 eV opens at $K_1$ and $K_2$, inducing a nonvanishing Berry curvature. Fig.~\ref{fig:2}(b) shows the Berry curvature in the two valence bands. For one valence band (green solid line) the Berry curvature is negative at $K_1$ and positive at $K_2$, while it is exactly the opposite for the other valence band (purple dashed line). Thus the Berry curvature integrated within each valley ($K_1$ and $K_2$) vanishes. The same qualitative behavior is observed for increased values of the asymmetric potential $V^{\text{asym}}_{AB}$ $=$ 0.01 eV larger than the backgate potential $V_{BG}$ $=$ 0.001 eV (Fig.~\ref{fig:2}(e),(f)) as well as asymmetric potential $V^{\text{asym}}_{AB}$ $=$ 0.01 eV smaller than the backgate potential $V_{BG}$ $=$ 0.05 eV (Fig.~\ref{fig:2}(i),(j)). In these cases the Berry curvature is spread out a bit more in momentum space. In the latter case with larger backgate potential one can observe additional band crossings to the left and right of the Dirac points (Fig.~\ref{fig:2}(i)) with vanishing Berry curvature (Fig.~\ref{fig:2}(j)).

\begin{figure*}[htp!]
    \centering
    \includegraphics[width=1.0\linewidth]{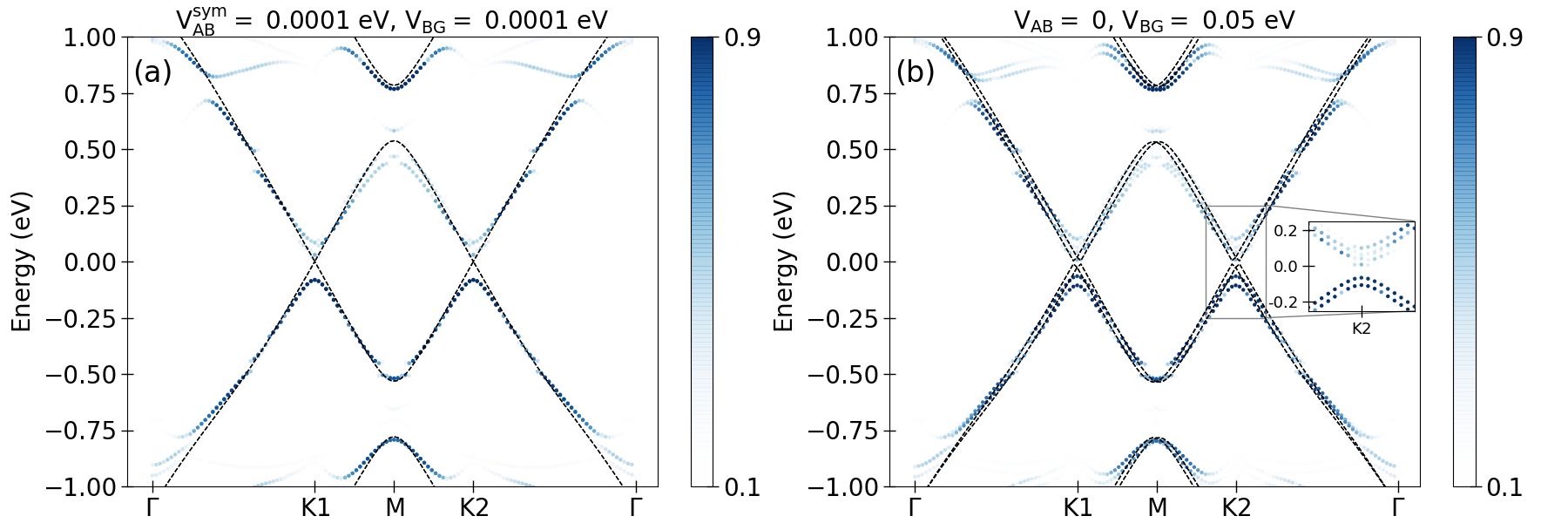}
    \caption{Floquet band structures for a strong circularly polarized driving field with ${E_{max}=13.7}$~MV/cm ($A=0.15$~$\text{\AA}^{-1}$ at $\Omega = 2.23$ eV)
    for different onsite potentials ${V_{AB}}$ and backgate voltages ${V_{BG}}$. 
    The black dashed lines indicate the equilibrium bands. The overlap between the Floquet bands and the equilibrium bare bands is denoted by the blue color as indicated in the color bars. In both cases the breaking of time-reversal symmetry adds mass to the Dirac bands, which results in a finite energy gap at the Dirac points. Additionally, gaps open away from the Dirac points, due to resonant coupling between Floquet bands. The inset in (b) shows a close-up of the region near $K_2$.}
    \label{fig:3} 
\end{figure*}  

We now study what happens for the case of a symmetric $A$-$B$ potential $V^{\text{sym}}_{AB}$. For infinitesimally small potentials $V^{\text{sym}}_{AB}$ $=$ $V_{BG}$ $=$ 0.0001 eV (Fig.~\ref{fig:2}(c),(d)) both valence bands show the same Berry curvature being positive near $K_1$ and negative near $K_2$. Thus in the symmetric case, as opposed to the asymmetric case, there is a non-vanishing total valley Berry curvature but vanishing overall Berry curvature due to cancellation between both valleys in the mBZ. 
The same qualitative behavior is observed for increased values of the symmetric potential $V^{\text{sym}}_{AB}$ $=$ 0.01 eV larger than the backgate potential $V_{BG}$ $=$ 0.001 eV (Fig.~\ref{fig:2}(g),(h)). This behavior is switched when the backgate potential $V_{BG}$ $=$ 0.05 eV is larger than the symmetric potential $V^{\text{asym}}_{AB}$ $=$ 0.01 eV (Fig.~\ref{fig:2}(k),(l)). In this case the Berry curvature is analogous to the one for the asymmetric case with vanishing valley Berry curvature for each valley individually. This finding is particularly interesting as it opens up the possibility of starting from a topologically trivial insulating phase for $V_{BG}$ $\neq$ 0 and $V_{AB}$ $=$ 0, which is the realistic scenario for TBG. Below we will show that by applying circularly polarized light, one can light-induce the transition from topologically trivial to topologically nontrivial insulating band structures at a nonvanishing critical field strength, which is not possible for single-layer graphene.

Overall we note that TBG obviously offers more possibilities for combinations of Berry curvature than single-layer graphene. In particular the Berry curvature can vanish per valley when summed over all valence bands, while being nonzero for each band separately within a valley, for broken inversion symmetry due to an asymmetric $A$-$B$ potential or due to a symmetric $A$-$B$ potential when the backgate voltage exceeds this $A$-$B$ potential. This scenario obviously does not exist for single-layer graphene which has only one valence band (per spin, which is suppressed here) in the Dirac bands. In all cases discussed here the total Berry curvature and thus the valence band Chern number vanishes when time-reversal symmetry is intact, like in single-layer graphene or the Haldane model with time-reversal symmetry. 
Importantly, we do not make a claim about the topology of the regime at and below the highest magic angle around 1.05$^\circ$, for which there have been several predictions for the effective Berryology \cite{po_origin_2018, zou_band_2018, de_gail_topologically_2011, koshino_maximally_2018}.

\begin{figure*}[htp!]
    \centering
    \includegraphics[width=1.0\linewidth]{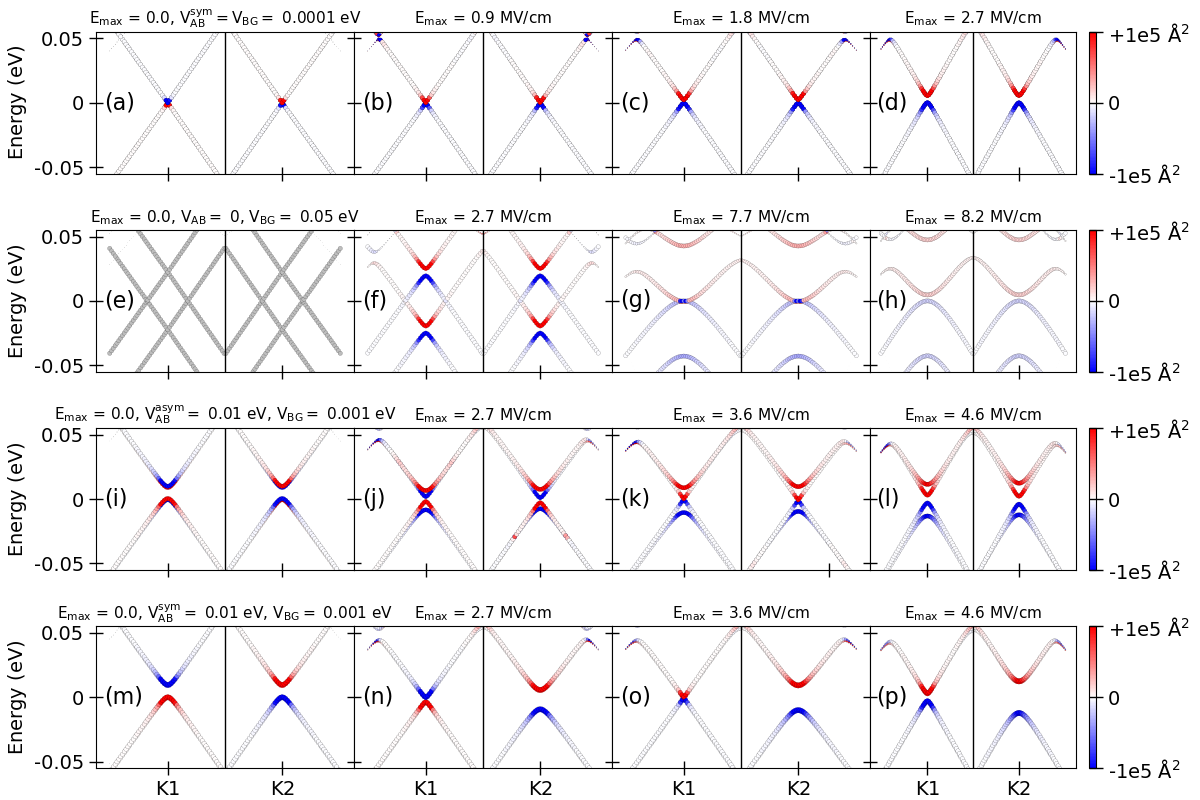}
    \caption{Floquet-engineered Berry curvature near the two Dirac points of the mBZ along {$\Gamma$-$K_1$-$M$-$K_2$-$\Gamma$} for different onsite potentials $V_{AB}$, backgate voltages $V_{BG}$, and increasing circularly polarized field amplitudes $E_{max}$. The size of the points indicate the overlap with the original equilibrium bands. Throughout, ${\Delta_\Gamma=2.23}$~eV is chosen as Floquet driving frequency. (a),(e),(i),(m) Equilibrium case. In (e) the Berry curvature is not defined at $K_1$ ($K_2$), due to band degeneracy, and zero otherwise. (b) For infinitesimally small potentials, a finite driving amplitude immediately opens a gap and induces a nontrivial Floquet-band topology. (f),(i),(j),(m),(n) Topologically trivial phase, characterized by zero integrated Berry curvature of the valence (conduction) bands. (g),(k),(o) Topological transition indicated by gap closing. (g) Trivial shift of bands while the sign of associated Berry curvature is preserved. (k) Gap closing and flow of Berry curvature between lowermost conduction band and uppermost valence band. (o) Gap closing and flow of Berry curvature both between bottom conduction band and valence band and between top conduction band and valence band, respectively. (b),(c),(d),(h),(i),(p) Chern-insulating phase, characterized by re-opened gaps and a finite integrated Berry curvature in the valence (conduction) bands.}
    \label{fig:4} 
\end{figure*}    

\begin{figure*}[htp!]
    \centering
    \includegraphics[width=0.9\linewidth]{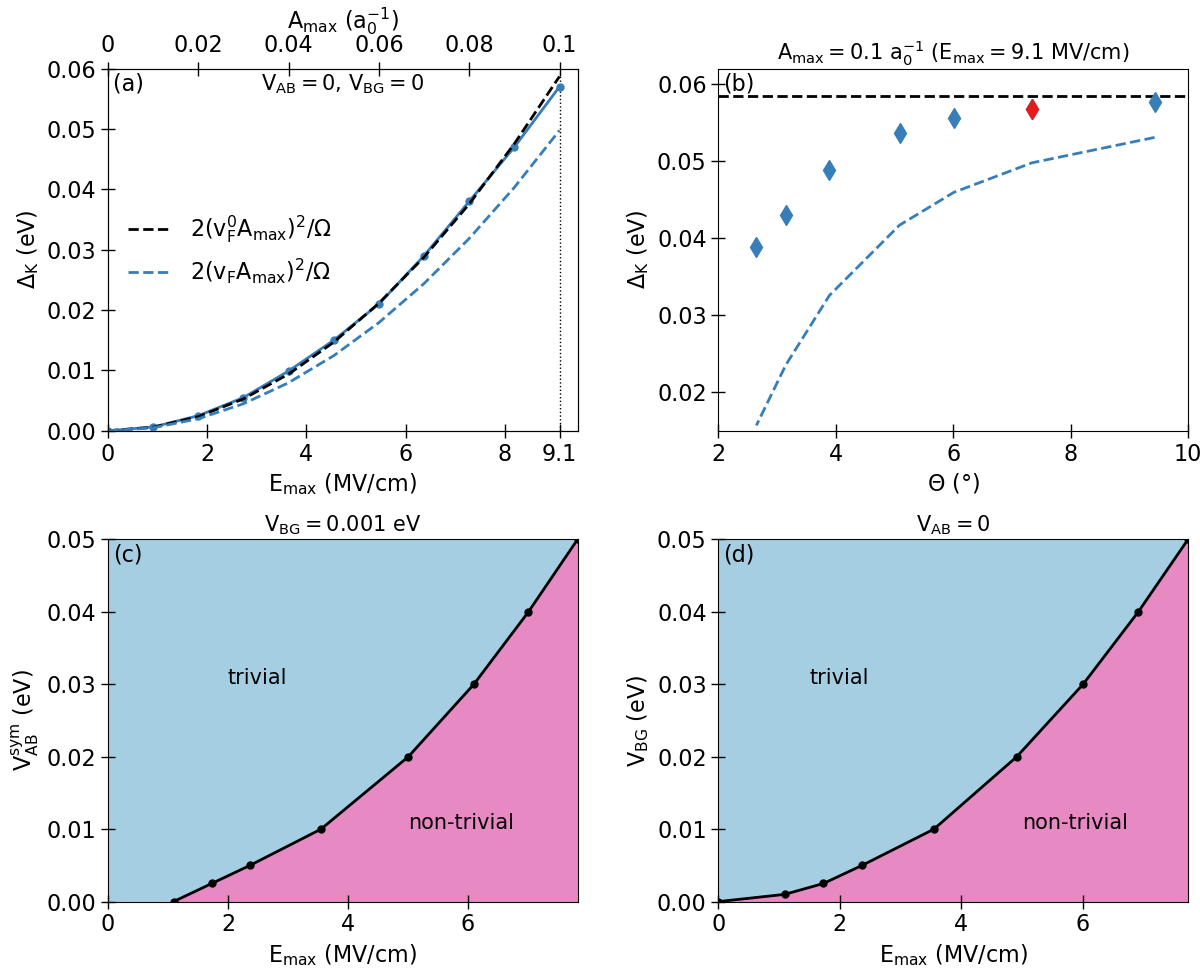}
    \caption{Floquet topological gaps and phase diagrams. (a) Energy gap at $K$ as a function of the driving field amplitude. The blue (black) dashed line shows the quadratic dependence of the gap on the driving amplitude, $A_{max}$, and the equilibrium TBG (monolayer graphene) Fermi velocity, $v_F$ ($v_F^0$), in the low-amplitude, high-frequency limit. The black dotted line indicates the field strength chosen in panel (b). (b) Energy gap as a function of twist angle for a fixed field strength. The value used throughout the paper, $\Theta=7.34^\circ$, is indicated by red color. The black dashed line indicates the corresponding single-layer Floquet band gap. The blue dashed line shows the gap values which would be naively expected from the renormalized Fermi velocity. (c),(d) The pink (light blue) area indicates the topologically nontrivial (trivial) phase. (c) With symmetric sublattice potential and small backgate voltage. Due to the additional backgate voltage, the topologically nontrivial phase always requires nonzero driving amplitude. (d) With backgate voltage only. Throughout, a photon frequency, $\Omega$ $=$ 2.23 eV, is assumed.} 
    \label{fig:5}
\end{figure*}

\section{Floquet band structure}\label{Sec.III}

We now turn to the case of laser-driven TBG with circularly polarized light in which Floquet bands are formed. Focusing again on a twist angle ${\Theta = 7.34^\circ}$, used here and in all that follows, we choose a photon frequency $\Omega =$ 2.23 eV tuned to the bandwidth $\Delta_{\Gamma}$ for this particular twist angle. This choice is motivated by the fact that Floquet sidebands overlapping with the original bands are not created in this case within the low-energy Dirac bands, but only with higher lying bands. The laser field is described by a time-dependent, spatially homogeneous vector potential ${\bm{A}(t) = A_{\text{max}}(\bm{e}_x \sin(\Omega t) + \bm{e}_y \cos(\Omega t))}$, and is coupled to the electrons in TBG via Peierls substitution,
\begin{equation}
    t_{ij} \rightarrow{} \tilde{t}_{ij}(t) = t_{ij}e^{\I \bm{A}(t)\bm{r}_{ij}}.  
\end{equation}
From the resulting time-dependent Hamiltonian we compute the Floquet bands by expanding in Floquet harmonics \cite{shirley_solution_1965,sambe_steady_1973,eckardt_high-frequency_2015} and diagonalizing the Floquet Hamiltonian with a cutoff in the Floquet index and checked convergence in this cutoff. In order to plot the resulting bands and their topology we choose to project the Floquet bands onto the original energy eigenstates of the bare Hamiltonian, i.e., to compute the square of the wavefunction overlap and color the resulting Floquet bands with a color scale according to this overlap (see Sec.~\ref{app:floquet}). 

In Fig.~\ref{fig:3} we show the original equilibrium bands (dashed lines) and the projected Floquet bands for a relatively strong laser field (${E_{max}=13.7}$~MV/cm) and two cases, the one of infinitesimal potentials $V^{\text{sym}}_{AB}$ $=$ $V_{BG}$ $=$ 0.0001 eV (Fig.~\ref{fig:3}(a)) and the one of vanishing A-B potential but nonvanishing backgate potential $V_{BG}$ $=$ 0.05 eV (Fig.~\ref{fig:3}(b)). For the case in Fig.~\ref{fig:3}(a) the bare bands correspond essentially to the Dirac bands of pristine TBG. The circularly polarized driving field opens a sizeable energy gap at both Dirac points $K_1$ and $K_2$ and creates a variety of Floquet bands, most of them with relatively small overlap with the equilibrium bands. Nevertheless a number of avoided band crossings are visible in the chosen energy window, leading to a more complicated structure than for single-layer graphene. Similarly, for the case in Fig.~\ref{fig:3}(b) the main feature induced by the laser driving is a band gap opening at $K_1$ and $K_2$ but now with visibly split Floquet valence and conduction bands due to the sizeable backgate potential. 

\section{Floquet-engineered topology}\label{Sec.IV}
Finally, the obvious follow-up question to address is the one of the light-induced Berry curvature and corresponding Floquet-band topology. To this end we focus on the low-energy Floquet-Dirac bands near $K_1$ and $K_2$ and color these bands according to their respective Berry curvatures. The resulting Floquet-engineered Berry curvatures are presented in Fig.~\ref{fig:4}. Fig.~\ref{fig:4}(a)-(d) shows the evolution for infinitesimal potentials $V^{\text{sym}}_{AB}$ $=$ $V_{BG}$ $=$ 0.0001 eV as a function of increasing driving field strength $E_{\text{max}}$. Initially in equilibrium (Fig.~\ref{fig:4}(a)) one observes a nonzero Berry curvature in each valley but with flipped sign between valleys (cf.~Fig.~\ref{fig:2}(d)). This immediately changes when a driving field is applied (Fig.~\ref{fig:4}(b)), which leads to a band inversion and sign change of the Berry curvature around $K_1$ and a non-vanishing total Berry curvature in the valence bands (conduction bands) due to broken time-reversal symmetry exceeding the breaking of inversion symmetry. Similarly to the Haldane model case in the high-frequency driving limit, this is then expected to lead to an almost quantized light-induced anomalous Hall effect but with an additional double degeneracy due to the bilayer instead of single-layer structure (4$e^2/h$ instead of 2$e^2/h$). The observed band gap and spreading of the Berry curvature in momentum space become more pronounced as the field strength increases (Fig.~\ref{fig:4}(c),(d)).

We now turn to the case of vanishing A-B potential and nonzero backgate potential $V_{BG}$ $=$ 0.05 eV (Fig.~\ref{fig:4}(e)-(h)). Initially the Berry curvature vanishes identically everywhere (Fig.~\ref{fig:4}(e)). Turning on the laser field, a nonvanishing Berry curvature is induced but with compensation between the two valence bands (conduction bands) for a moderate field (Fig.~\ref{fig:4}(f)). Increasing the field strength more one reaches a critical field (Fig.~\ref{fig:4}(g)) at which the top valence band and bottom conduction band invert and the net valence band Berry curvature as well as the net conduction band Berry curvature become nonzero. This topological phase transition is completed at even larger field strength (Fig.~\ref{fig:4}(h)).

A topological Floquet-engineered phase transition is induced at smaller field strength when the backgate voltage is smaller, $V_{BG}$ $=$ 0.001 eV (Fig.~\ref{fig:4}(i)-(l)), where we also include an asymmetric $A$-$B$ potential $V^{\text{asym}}_{AB}$ $=$ 0.01 eV larger than the backgate potential. This is the same as the scenario discussed before where the valley Berry curvature vanishes when integrated over all valence bands in the absence of the laser field. This is changed by the laser driving and a band inversion and flow of Berry curvature between top valence and bottom conduction bands is observed simultaneously in both valleys at a critical field strength (Fig.~\ref{fig:4}(k)). 

By contrast, for $V_{BG}$ $=$ 0.001 eV and a larger symmetric onsite potential $V^{\text{sym}}_{AB}$ $=$ 0.01 eV (Fig.~\ref{fig:4}(m)-(p)), we have a net nonzero valley Berry curvature (Fig.~\ref{fig:4}(m)), qualitatively equivalently to the case of the single-layer Haldane model. In this case a moderate driving field creates asymmetric energy gaps at $K_1$ versus $K_2$ (Fig.~\ref{fig:4}(n)), where one of the gaps at $K_2$ increases due to the laser, while the other one at $K_1$ decreases. A gap closing and band inversion is then observed at $K_1$ for a critical field strength (Fig.~\ref{fig:4}(o)) with a completed transition to a topological Chern-insulating Floquet band structure above the critical field strength ((Fig.~\ref{fig:4}(p))).

We note that these Floquet-engineered Berry curvatures could be measured in a momentum-, energy- and time-resolved fashion using circularly polarized time- and angle-resolved photoemission spectroscopy \cite{schuler_local_2019}. 
Finally, we summarize our findings for the Floquet-engineered band gaps and topology in Fig.~\ref{fig:5}. Fig.~\ref{fig:5}(a) shows the Dirac point band gap at zero onsite potentials (pristine TBG) as a function of laser driving field strength, or as a function of peak vector potential, respectively. The blue (black) dashed line shows $2(v_F A_{\text{max}})^2/\Omega$ ($2(v^0_F A_{\text{max}})^2/\Omega$), respectively, with $v_F$ the renormalized Fermi velocity of TBG and $v^0_F$ the bare Fermi velocity of single-layer graphene. Interestingly, the light-induced band gap scales essentially with the bare Fermi velocity of single-layer graphene as opposed to the renormalized one in TBG (see Appendix for an extended discussion). For small twist angles, however, deviations from scaling with $v^0_F$ appear. The origin of this scaling behavior presumably lies in the fact that the renormalization of the Fermi velocity stems from interlayer hoppings, which are predominantly perpendicular to the plane of the individual graphene layers. In contrast, the light field is chosen to have in-plane polarization, thus not coupling to the out-of-plane currents. Therefore, the single-layer Fermi velocity is the key ingredient for the light-induced band gap.

Fig.~\ref{fig:5}(b) shows the Floquet band gap at the Dirac points for vanishing onsite potentials at different intermediate twist angles. The peak electric field strength, $E_{\text{max}}=9.1$~MV/cm, and the driving frequency, $\Omega= 2.23$~eV, are kept constant. The gap increases as a function of twist angle at small twist angles and then reaches a plateau approaching the gap value for single-layer graphene. Clearly, for $7.34^\circ$ the size of the gap is very close to the single-layer value. \footnote{The decreasing size of the light-induced gap due to increasing band flatness (smaller twist angles) is confirmed in Fig.~\ref{fig:FIG4_APP} of the Appendix, in which we show the gap dependency on the driving amplitude for two smaller angles.}

Fig.~\ref{fig:5}(c) shows the Floquet phase diagram with the trivial phase (vanishing valence band integrated Berry curvature) and nontrivial phases (non-zero integrated Berry curvature) as a function of field strength for varying symmetric $A$-$B$ potential and a small backgate voltage. Due to the backgate voltage, as seen already in Fig.~\ref{fig:4}, a finite critical field strength is required to switch the system to the topologically nontrivial phase. 
This changes in the case of a vanishing  $A$-$B$ potential and a finite backgate voltage, shown in Fig.~\ref{fig:5}(d). Here, essentially the shape of the phase boundary exhibits the same quadratic dependence on field strength as the energy gap in Fig.~\ref{fig:5}(a).

\section*{CONCLUSION AND OUTLOOK}\label{CONC}
We have shown how the Berry curvature in twisted bilayer graphene is affected by different types of inversion-symmetry breaking. The effects of circularly polarized light on the full tight-binding model of twisted bilayer graphene are found to be essentially as expected for two copies of single-layer graphene. Importantly, the light-induced energy gap is not strongly affected by interlayer coupling at twist angles much larger than the largest magic angle. Moreover, the opportunity to break inversion symmetry by backgating allows to tune the phase transition between topologically trivial and nontrivial states, in distinct contrast with single-layer graphene. This opens up the realistic prospect of a finite-field topological transition in solid-state experiments. 
A future task is to study theoretically the ultrafast light-induced transport properties of TBG for realistic driving pulses especially in the mid-infrared and for smaller twisting angles near 1.7$^\circ$, including the effects of excitation and dissipation in real time, which is a much more formidable task for TBG compared to a single branch of Dirac fermions \cite{sato_microscopic_2019, sato_light-induced_2019}. In particular, real-time signatures of tunneling into topological edge states might provide additional insights \cite{tuovinen_distinguishing_2019}. An intriguing opportunity opens up for small twist angles, at which the distance between both Dirac points in the mini-Brillouin zone becomes very small. In this case, the laser field strength could realistically lead to a peak vector potential that exceeds the distance $|K_2-K_1|$. In this case the band gap induced initially by a moderate field amplitude could close again leading to a second topological phase transition, which should lead to a signature in the Hall transport experiment. Moreover it is also interesting to address the question of light-induced topology without classical driving fields by encapsulating TBG in a cavity, as recently proposed for single-layer graphene \cite{wang_cavity_2019}. 

Finally the possibility to not only tune single-particle bandwidth and Berry curvature but also interaction effects \cite{grushin_floquet_2014,wolf_electrically-tunable_2019} could pave the way for solid-state platforms with exotic nonequilibrium phases of matter analogous to Floquet-engineered interaction phases in artificial lattices \cite{aidelsburger_measuring_2015}. Moreover, the concept of twistronics can also be extended to other two-dimensional materials, such as hexagonal boron nitride \cite{xian_multi-flat_2018} and GeSe\cite{kennes_new_2019}, and it is an intriguing subject for future study to investigate Floquet engineering in these materials.

\section*{ACKNOWLEDGEMENT} 
Discussions with S.~A.~Sato are gratefully acknowledged. G.E.T.~and M.A.S.~acknowledge financial support by the DFG through the Emmy Noether program (SE 2558/2-1). This work was supported by the European Research Council (ERC-2015-AdG694097).
The Flatiron Institute is a division of the Simons Foundation.

\bibliography{TBG_Floquet}

\clearpage
\appendix
\section{Model} \label{app:model}
\begin{figure}[htp!]
    \centering
    \includegraphics[width=1.0\linewidth]{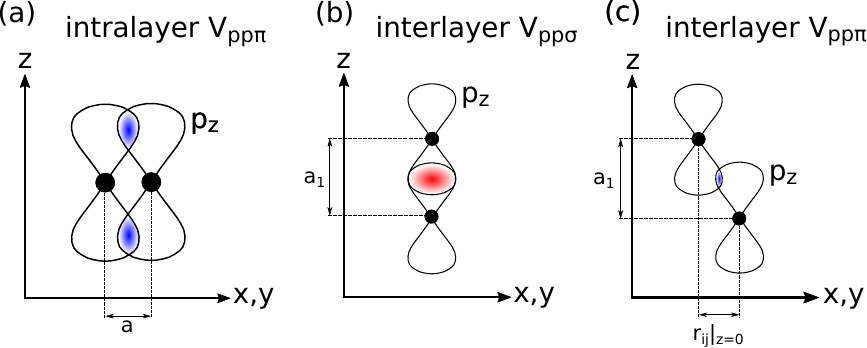}
    \caption{Tight-binding hopping elements. Only $p_z$-orbitals are taken into account. (a) Intralayer hopping is restricted to $pp\pi$ terms. There are two channels for interlayer hopping: Vertical hopping is dominated by the $pp\sigma$ term (b). For non-vertical transitions also $pp\pi$ terms (c) contribute.}
    \label{fig:FIG1_APP}
\end{figure}

\begin{figure}[htp!]
    \centering
    \includegraphics[width=1.0\linewidth]{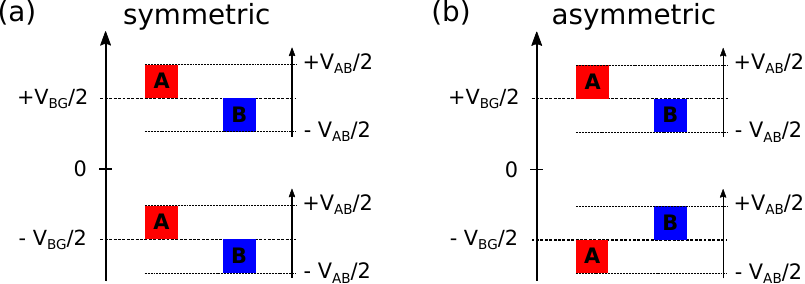}
    \caption{Onsite potentials that break inversion symmetry. The backgate voltage, $V_{BG}$, is always chosen symmetric between both layers. (a) Symmetric $\pm/\pm$ sublattice potential, $V_{AB}^{sym}$. (b) Asymmetric $\pm/\mp$ sublattice potential, $V_{AB}^{asym}$.}
    \label{fig:FIG2_APP}
\end{figure}

The tight-binding Hamiltonian is of the general form
\begin{equation}
    H = \sum_{i}\epsilon_{i}c_{i}^\dagger c_{i} + \sum_{i\neq j}t_{ij} c_{i}^\dagger  c_{j}, 
\end{equation}
where $c_i^\dagger$ ($c_i$) creates (annihilates) an electron in the $p_z$ orbital at the atomic position $\bm{r}_i$. We use hopping matrix elements as introduced in \cite{trambly_de_laissardiere_numerical_2012},
\begin{equation}
    t_{ij} = n^2 V_{pp\sigma}(r_{ij}) +(1-n^2)V_{pp\pi}(r_{ij}),
\end{equation}
where ${r_{ij}=||\bm{r}_{ij}||}$ and ${n=z_{ij}/r_{ij}}$. For intralayer hopping, only the $V_{pp\pi}$ term contributes, as ${z_{ij}=0}$. For interlayer hopping, both the $V_{pp\pi}$ and the $V_{pp\sigma}$ terms are nonzero (${z_{ij}=a_1}$). A detailed illustration is shown in Fig.~\ref{fig:FIG1_APP}. Both hopping contributions are assumed to decay exponentially as a function of distance,
\begin{eqnarray}
    V_{pp\pi} = \gamma_0 \exp{[q_\pi(1-r_{ij/a})]}, \nonumber \\
    V_{pp\sigma} = \gamma_1 \exp{[q_\sigma(1-r_{ij/a_1})]}.
\end{eqnarray}
Here, ${a=1.418}$~\AA is the intralayer nearest-neighbor distance, and ${a_1 = 3.364}$~\AA is the interlayer distance. The monolayer lattice constant of the triangular Bravais lattice is ${a_0=2.445}$~\AA. For the nearest-neighbor hopping, ${\gamma_0 = -3.24}$~eV, we add $20\%$ to the DFT-fitted value in order to compensate for many-body effects beyond the employed DFT functional and in order to fit the van-Hove singularity in the tunneling density of states reported in \onlinecite{kerelsky_magic_2018}. For the $\sigma$-related hopping we find ${\gamma_1 = 0.55}$~eV. For the exponential decay we find fitting parameters, $q_\pi=3.15$ and $q_\sigma= q_\pi\frac{a_1}{a}$. 
For the fictitious local potentials on the $A$ and $B$ sublattices that break inversion symmetry, $\epsilon_i$, we assume different symmetry configurations, as shown in Fig.~\ref{fig:FIG2_APP}. We assume periodic Born-von Karman boundary conditions within in the $x$-$y$-plane. The Hamiltonian in momentum space is then calculated by a Fourier transforms, ${c^\dagger(\bm{k})=\sqrt{V_{\text{BZ}}^{-1}}\sum_i c_i^\dagger \exp(\I\bm{k}\bm{r}_i)}$ and ${c(\bm{k})=\sqrt{V_{\text{BZ}}^{-1}}\sum_i c_i \exp(-\I\bm{k}\bm{r}_i)}$, where $V_\text{BZ}$ is the volume of the two-dimensional Brillouin zone.

\section{Floquet dynamics} \label{app:floquet}
A circularly polarized laser driving field is included by coupling to a time-dependent external gauge field, ${\bm{A}(t) = A_{\text{max}}(\bm{e}_x \sin(\Omega t) + \bm{e}_y \cos(\Omega t))}$, where $\bm{e}_x$ and $\bm{e}_y$ denote the unit vectors in the spatial in-plane directions, via Peierls substitution of the form
\begin{equation}
    t_{ij} \rightarrow{} \tilde{t}_{ij}(t) = t_{ij}e^{\I \bm{A}(t)\bm{r}_{ij}}.  
\end{equation}
The main effect of the circularly polarized field is the breaking of time-reversal symmetry which induces a mass term to the Dirac fermions, as has been investigated for single-layer graphene \cite{oka_photovoltaic_2009,
sentef_theory_2015}. The peak electric field strength is calculated from the amplitude of the vector potential by the relation, ${E_{\text{max}} = \frac{A_{\text{max}}}{a_0}\Omega}$,  where ${a_0=2.445}$~\AA, is the monolayer lattice constant. The resulting non-equilibrium state of the light-matter coupled system is periodic in time and can thus be analyzed by Floquet theory \cite{shirley_solution_1965,sambe_steady_1973,eckardt_high-frequency_2015}, which, by a Floquet-Bloch decomposition, maps the bare time-dependent Hamiltonian, ${H(t+T)=H(t)}$, to the time-independent Floquet Hamiltonian,
\begin{eqnarray}
    \mathcal{H}^{nm} = \frac{1}{T} \int_T \D t e^{\I (m-n)\Omega t} H(t) + \delta_{mn} m \Omega 1\!\!1. \label{eq:FLOQUET_HAM}
\end{eqnarray}
The integers $m$ and $n$ arise from the Floquet expansion of the time-dependent solutions of the Schroedinger equation,
\begin{equation}
    \ket{\Psi(t)} = \sum \limits_{m=-\infty}^{+\infty} \exp[-\I(\epsilon+m\Omega)t]\ket{u_m},
\end{equation}
where $\epsilon$ is the Floquet quasi-energy and $\ket{u_m}$ the corresponding Floquet eigenfunction of the Floquet Hamiltonian. For the results shown in the main text we truncate the expansion after order ${|m|=2}$ and use 200 sampling points to perform the integral over one period ${T=2\pi/\Omega}$ in Eq.~\ref{eq:FLOQUET_HAM} numerically with the trapezoidal rule. We ensure convergence in the Floquet cutoff by means of the energy gap at the Dirac points with error smaller than $10^{-5}~\text{eV}$. In order to highlight the  bands which overlap with the bare bands of the undriven system, in contrast to Floquet sidebands, we calculate the squared overlap, $\sum_\alpha |\langle \alpha | e_j \rangle |^2$, of the Floquet energy eigenstates, $|e_j \rangle$, with the energy eigenstates of the bare Hamiltonian, $H| \alpha \rangle = E_\alpha | \alpha \rangle$. Throughout this work we use units where ${e=\hbar=c=1}$. 

\section{Berry curvature} \label{app:curvature}
\begin{figure}[htp!]
    \centering
    \includegraphics[width=1.0\linewidth]{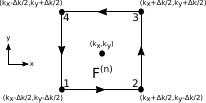}
    \caption{Calculation of the Berry flux. The Berry flux of the $n$-th band, $F^{(n)}$, for a quasimomentum $\bm{k}=(k_x,k_y)$ is calculated by a closed loop along the eigenstates at position 1 to 4 in momentum space. Throughout we use a grid spacing of $\Delta k = 10^{-5}$~$\text{\AA}^{-1}$.}
    \label{fig:FIG3_APP}
\end{figure}

The Berry curvature at a fixed momentum point can be defined by the Berry flux of an infinitesimally small loop around that point, divided by the enclosed area \cite{asboth_short_2016},
\begin{equation}
    B^{(n)} = \lim_{\Delta k_x, \Delta k_y \to 0} \frac{F^{(n)}}{\Delta k_x, \Delta k_y},
\end{equation}
where $n$ refers to the corresponding band. On a discretized grid, the flux can be written in the form \cite{resta_theory_2007}
\begin{equation}
    F^{(n)} \approx \text{Im} \left( \ln\prod \limits_{j}\left \langle u_{n,\bm{k}_j} | u_{n,\bm{k}_{j+1}} \right \rangle  \right).
\end{equation}
Here, $u_{n,\bm{k}_i}$ refers to the $n$-th energy eigenstate of the bare Hamiltonian and the Floquet Hamiltonian, respectively, at quasi-momentum $\bm{k}_i$, where ${i \in \{1,2,3,4\}}$, as depicted in Fig.~\ref{fig:FIG3_APP}. We use a grid spacing of $\Delta k = 10^{-5}$~$\text{\AA}^{-1}$ for which the Berry curvature is sufficiently converged.

\section{Light-induced energy gaps}

\begin{figure}[htp!]
    \centering
    \includegraphics[width=1.0\linewidth]{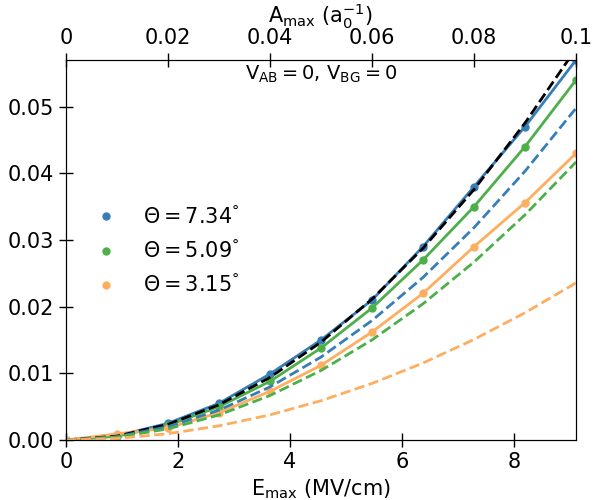}
    \caption{Full energy gap at the Dirac point as a function of the driving amplitude for different twist angles, $\Theta$ (colored dots). In all three cases, the Floquet driving frequency is $\Omega=2.23$~eV. The black dashed line indicates the monolayer dependence of the gap on the driving amplitude and the Fermi velocity, $2(v_F^0 A_{\text{max}})^2/\Omega$. The colored dashed lines indicate the corresponding relation for the twisted case.}
    \label{fig:FIG4_APP}
\end{figure}

In Fig.~\ref{fig:FIG4_APP} we show the light-induced energy gaps at the Dirac point without inversion-symmetry breaking fields for different twist angles as indicated. In all three cases, we choose as the photon frequency the $\Gamma$-point low-energy bandwidth, $\Delta_\Gamma =2.23$~eV, which we extract from the equilibrium band structure for $\Theta=7.34^\circ$. Importantly we find almost perfect scaling behavior with the bare Fermi velocity of single-layer graphene, as opposed to the actual Fermi velocity of the bilayer, for the largest twist angles of 7.34$^{\circ}$. For smaller twist angles we find a deviation and the actual gap lies in between the one expected for single-layer graphene and the one for the bilayer. 

\end{document}